\documentclass[twocolumn,tighten]{aastex631}
\usepackage{amsfonts,amsmath,amssymb}
\usepackage{bm}
\usepackage{xcolor}
\usepackage[]{cleveref}%capitalise, noabbrev
\crefname{section}{\S\!}{\S\S\!}
\crefname{appendix}{\S\!}{\S\S\!}
\crefname{equation}{Eq.}{Eqs.}
\Crefname{equation}{Equation}{Equations}
\crefname{figure}{Fig.}{Figs.}
\Crefname{figure}{Figure}{Figures}
\usepackage{mathtools,upgreek,pifont}
\usepackage{cuted}

\def\apjs{ApJS}
\def\apj{ApJ}
\def\apjl{ApJL}

\def\aap{A\&A}

\def\mnras{MNRAS}

\def\prl{PhRvL}
\def\prx{PhRvX}
\def\jgr{JGR}

\def\ssr{SSRv}

\def\pop{PhPl}
\def\pof{PhFl}

\def\jcomp{JCoPh}
\def\jpp{JPlPh}

\def\solphys{SoPh}
\def\aapr{A\&ARv}
\def\natas{NatAs}
\def\jats{JAtS}
\def\jgrsp{JGRA}
\def\jcomp{JCoPh}
\def\lrsp{LRSP}
\def\njp{NJPh}

\usepackage{hyperref}
\definecolor{darkblue}{rgb}{0.0,0.0,0.3}
\hypersetup{colorlinks,breaklinks,
            linkcolor=darkblue,urlcolor=darkblue,
            anchorcolor=darkblue,citecolor=darkblue}

\newcommand{\pD}[2]{\frac{\partial #2}{\partial #1}}
\newcommand{\D}[2]{\frac{{\rm d} #2}{{\rm d} #1}}
\newcommand\bb[1]{\mbox{\boldmath{$#1$}}}
\newcommand\grad{\bb{\nabla}}
\newcommand\bcdot{\,\bb{\cdot}\,}
\newcommand\btimes{\,\bb{\times}\,}
\newcommand{\msb}[1]{\bb{\mathsf{#1}}}

\newcommand{\const}{{\rm const.}}
\newcommand{\ez}{\hat{\bb{z}}}

\newcommand{\rmd}{{\rm d}}

\newcommand{\vth}[1]{v_{{\rm th}{#1}}}

\shorttitle{Extreme heating of minor ions}
\shortauthors{Zhang et al.}

\begin{document}

\title[]{Extreme heating of minor ions in imbalanced solar-wind turbulence}
\author[0000-0002-3987-5977]{Michael F.~Zhang}
\affiliation{Department of Astrophysical Sciences, 
Princeton University, Peyton Hall, Princeton, NJ 08544, USA}
\affiliation{Princeton Plasma Physics Laboratory, PO~Box 451, Princeton, NJ 08543, USA}
\email{E-mail for correspondence: mfzhang@princeton.edu}
\author[0000-0003-1676-6126]{Matthew W.~Kunz}
\affiliation{Department of Astrophysical Sciences, 
Princeton University, Peyton Hall, Princeton, NJ 08544, USA}
\affiliation{Princeton Plasma Physics Laboratory, PO~Box 451, Princeton, NJ 08543, USA}
\author[0000-0001-8479-962X]{Jonathan Squire}
\affiliation{Physics Department, University of Otago, 730 Cumberland St, Dunedin 9016, New Zealand}
\author[0000-0001-6038-1923]{Kristopher G.~Klein}
\affiliation{Lunar and Planetary Laboratory, University of Arizona, Tucson, AZ 85721, USA}

\begin{abstract}
Minor ions in the solar corona are heated to extreme temperatures, far in excess of those of the electrons and protons that comprise the bulk of the plasma. These highly non-thermal distributions make minor ions sensitive probes of the collisionless processes that heat the corona and power the solar wind. The recent discovery of the ``helicity barrier" offers a mechanism in which imbalanced Alfv\'enic turbulence in low-$\beta$ plasmas preferentially heats protons over electrons, generating high-frequency, proton-cyclotron-resonant fluctuations. We use the hybrid-kinetic particle-in-cell code {\tt Pegasus++} to drive imbalanced Alfv\'enic turbulence in a 3D low-$\beta$ plasma with additional passive ion species, He$^{2+}$ and O$^{5+}$. A helicity barrier naturally develops, followed by clear phase-space signatures of oblique proton-cyclotron-wave heating and Landau-resonant heating from the imbalanced Alfv\'enic fluctuations. The former results in characteristically arced ion velocity distribution functions, whose non-bi-Maxwellian features are shown by linear {\tt ALPS} calculations to be critical to the heating process. Additional features include a steep transition-range electromagnetic spectrum, proton-cyclotron waves propagating in the direction of the imbalance, significantly enhanced proton-to-electron heating ratios, ion temperatures that are considerably more perpendicular with respect to magnetic field, and extreme heating of heavier species in a manner consistent with mass scalings inferred from spacecraft measurements. None of these features are realized in an otherwise equivalent simulation of balanced turbulence. If seen simultaneously in the fast solar wind, these signatures of the helicity barrier would testify to the necessity of incorporating turbulence imbalance in a complete theory for the evolution of the solar wind.
\end{abstract}


%\maketitle

\section{Introduction}
\label{sec:introduction}

Observations of alphas (He$^{2+}$) and heavier (``minor'') ions in the fast solar wind reveal extreme species-dependent temperatures that are far greater than that of the bulk protons \citep{Verscharen19,Tracy2015}. For example, remote sensing UVCS observations find O$^{5+}$ temperatures to be more than 20~times that of protons at around 2~solar radii \citep{Kohl2006}, and data taken by the {\em Wind} spacecraft near 1~au reveal alpha temperatures up to 6~times that of protons \citep{Kasper17}. More recently, in-situ measurements in the inner heliosphere taken by {\it Parker Solar Probe} (PSP) \citep{Mcmanus2024,Mostafavi24} and {\it Solar Orbiter} (SO) \citep{Bruno2024} reveal alpha temperatures 8 to 10 times the proton temperature.
These alpha and minor-ion temperatures also possess a strong anisotropy with respect to the mean magnetic field, with the field-perpendicular temperature exceeding the field-parallel temperature, $T_\perp > T_\parallel$ \citep{kohl98}.
Such differential energization of solar-wind particles provides valuable constraints on the processes that power the solar wind. This is particularly true for minor ions, whose low abundances (${\lesssim}0.1\%$) and variety of charges and masses make them well suited as test-particle tracers for local processes that occur throughout the solar wind and down to the lower corona where ion charge states are frozen in \citep{Bame1974}.

The processes that energize particles and continually power the solar wind are thought to derive their energy from the abundant Alfv\'enic turbulence that pervades the interplanetary medium \citep{Belcher71,Tu95,Chen2020}.
Stochastic heating is one such process, capable of perpendicularly heating ions preferentially at low values of the plasma $\beta$ parameter (the ratio of the thermal and magnetic pressures). In stochastic heating, ions undergo diffusion in perpendicular energy via a random walk caused by kicks from finite-amplitude turbulent fluctuations whose cross-field variation is comparable to the ion gyro-scale \citep{Chandran2010a,Cerri2021}. The degree to which ions undergo stochastic heating depends sensitively on the ratio of the electrostatic energy in the gyro-scale potential fluctuations and the thermal kinetic energy of the ions, with ratios ${\gtrsim}0.1$ leading to substantial ion energization. Given that heavier ion species have slower characteristic thermal velocities, and that such species' larger gyro-radii sample larger fluctuation amplitudes within the turbulent cascade, it is natural to expect stochastic heating to be greater for heavier species \citep{Chandran2013}.

Another process known to heat ions perpendicularly is resonant ion-cyclotron heating, for instance via ion-cyclotron waves (ICWs) \citep{Hollweg2002,Isenberg2011}. Models have demonstrated how a spectrum of ICWs can preferentially heat minor ions under solar-wind conditions \citep{Isenberg2007}. In-situ measurements taken by PSP show clear signatures of ICWs and ion-cyclotron heating occurring in the solar wind \citep{Bowen2020a,Vech2021,Bowen2022,Liu2023,Bowen2023,Bowen2024}, and these ICWs have been seen all the way out to 1~au by {\em Wind} \citep{Jian2009}. However, propagation of ICWs from the corona out to many solar radii is unlikely to be efficient \citep{Hollweg2000}, and PSP measurements of ICW-proton energy transfer rates suggest local generation of ICWs \citep{Vech2021,Bowen2024}. The process responsible for such local generation in the solar wind remains uncertain. In particular, balanced Alfv\'enic turbulence (where the energy in outgoing Alfv\'en-wave packets balances that of incoming Alfv\'en-wave packets, $|\bb{z}^{+} | \approx |\bb{z}^{-} |$ for the Els\"{a}sser variables $\bb{z}^{\pm}$) that undergoes a critically balanced cascade, such that the linear propagation time of an Alfv\'en-wave packet is of the same order as the non-linear interaction time ($\tau_{\rm A} \sim \tau_{\rm nl}$; \citealt{Goldreich1995}), generally produces highly anisotropic fluctuations with $k_{\parallel} \ll k_{\perp}$ \citep{Goldreich1995,Mallet2015}. Because of this strong anisotropy, it is difficult for Alfv\'enic fluctuations to attain the high frequencies required to produce ICWs before the turbulent cascade reaches the ion-Larmor scale, $k_{\perp} \rho_{i} \sim 1$, beyond which the cascading fluctuations become (dispersive and Landau-damped) kinetic AWs \citep{Schekochihin09,Howes08b}.

That being said, Alfv\'enic turbulence in the solar wind is often observed to be imbalanced, {\em viz.}, Alfv\'en-wave packets are observed to propagate dominantly in one direction over the other (e.g., $|\bb{z}^{+} | \gg |\bb{z}^{-} |$) such that the turbulence possesses a non-zero cross helicity, $\sigma_{\rm c}\doteq \langle |\bb{z}^{+}|^{2} - |\bb{z}^{-}|^{2} \rangle / \langle |\bb{z}^{+}|^{2} + |\bb{z}^{-}|^{2}\rangle\ne 0$ \citep{Roberts1987,Marsch2006,DAmicis2021}. 
Theoretical arguments based on gyrokinetic theory \citep{Schekochihin09,Meyrand2021} have shown that a generalized form of helicity is conserved in low-$\beta$ systems, which reverts to the cross helicity at scales above the proton-Larmor scale ($k_\perp\rho_{\rm p}\ll 1$) and becomes the magnetic helicity at scales below the proton-Larmor scale ($k_\perp\rho_{\rm p}\gg 1$). Combined with conservation of energy, the conservation of the cross helicity at $k_\perp\rho_{\rm p}\ll 1$ implies a forward cascade to smaller scales, while the conservation of the magnetic helicity at $k_\perp\rho_{\rm p}\gg 1$ implies an inverse cascade. The joint conservation of these two invariants prohibits the imbalanced portion of the injected energy from reaching sub-$\rho_{\rm p}$ scales, resulting in a ``helicity barrier'' that allows only the smaller, balanced portion of the flux to pass through to sub-ion-Larmor scales \citep{Meyrand2021}. The consequent build up of inertial-range fluctuation amplitudes above the helicity barrier results, via critical balance, in a decrease in the field-parallel scales of the fluctuations, and therefore an increase in their linear frequencies. Once those frequencies become comparable to the proton Larmor frequency, the turbulence excites proton-cyclotron waves (PCWs) that preferentially heat protons perpendicularly \citep{Squire2022,Squire2023}.

The recent discovery of the helicity barrier and consequent perpendicular heating of protons by PCWs, alongside ongoing measurements being taken by PSP and SO, make now an ideal time to revisit particle heating in imbalanced turbulence. Here we present and discuss results from two 3D hybrid-kinetic simulations designed to elucidate the heating of protons, alphas, and minor ions in Alfv\'enic turbulence that is relevant to the low-$\beta$ solar wind. The first simulation is essentially a multi-ion version of the run presented in \citet{Squire2023}, in which imbalanced Alfv\'enic turbulence was persistently driven in a collisionless, low-$\beta$, magnetized plasma. In this simulation, a helicity barrier naturally forms, which causes turbulent energy to accumulate in the inertial range until the critically balanced fluctuations attain ion-cyclotron frequencies and subsequently dissipate by heating the resonant ion species. The second simulation focuses instead on balanced turbulence and the consequent differential heating of ions, but is otherwise equivalent in terms of physical and numerical parameters \citep[cf.][]{Arzamasskiy2019,Cerri2021}.

The remainder of the article is organized as follows. Section~\ref{sec:equations} presents the model equations and how they are solved numerically to realize imbalanced (or balanced) Alfv\'enic turbulence of relevance to the solar wind. Section~\ref{sec:results} describes the time evolution of turbulence in both runs, with the imbalanced run exhibiting the formation of a helicity barrier and the subsequent excitation of oblique and parallel PCWs.  Unmistakable signatures of oblique-PCW heating and Landau damping manifest in the imbalanced-run velocity distribution functions (VDFs), which are found to be highly asymmetric and non-bi-Maxwellian. We solve the linear kinetic dispersion relation for these distinctive VDFs, with particular attention paid to the destabilization of parallel PCWs and the role they play in ion heating. The resulting extreme perpendicular temperatures of the ion species are found to be consistent with an approximately mass-proportional power-law scaling empirically derived from solar-wind observations \citep{Tracy2015,Tracy2016}. In contrast, no discernible helicity barrier or ion-cyclotron heating occurs in the balanced-but-otherwise-equivalent simulation. Instead, the VDFs are entirely consistent with theoretical predictions for stochastic heating  \citep{Chandran2013,KleinChandran2016}, following the analysis presented in \citet{Cerri2021}.

\section{Model equations and simulation approach}\label{sec:equations}

Our goal is to model Alfv\'enic turbulence in the collisionless solar wind, accounting for the finite-ion-Larmor-radius effects responsible for the formation of a helicity barrier and the kinetic effects required to achieve the barrier's subsequent regulation through the cyclotron-resonant heating of ions by excited PCWs. To do so, we adopt a hybrid-kinetic approach in which the ions are treated kinetically while the electrons constitute a neutralizing fluid \citep{byers78,hn78}. The distribution functions $f_i=f_i(t,\bb{r},\bb{v})$ of the various ion species $i$ are governed by the kinetic equation
\begin{equation}\label{eqn:vlasov}
    \pD{t}{f_i} + \bb{v}\bcdot\grad f_i + \frac{q_i}{m_i} \biggl(\bb{E} + \frac{\bb{v}}{c}\btimes\bb{B} \biggr)\bcdot\pD{\bb{v}}{f_i} = -\frac{q_i}{m_i} \bb{E}^{U}_{\rm ext}\bcdot\pD{\bb{v}}{f_i} ,
\end{equation}
where $q_i$ and $m_i$ are the charge and mass of species $i$, $c$ is the speed of light, $\bb{E}$ is the electric field, and $\bb{B}$ is the magnetic field. The latter evolves according to Faraday's law,
\begin{equation}\label{eqn:faraday}
    \frac{1}{c}\pD{t}{\bb{B}} + \grad\btimes\bb{E} = -\grad\btimes\bb{E}^{B}_{\rm ext} .
\end{equation}
The externally applied fields $\bb{E}^{U}_{\rm ext}$ and  $\bb{E}^{B}_{\rm ext}$ on the right-hand sides of Eqs.~\eqref{eqn:vlasov} and \eqref{eqn:faraday} are used to drive Alfv\'enic turbulence in the plasma; they are specified below. The electric field is obtained from the electron momentum equation,
\begin{equation}\label{eqn:emom}
    m_{\rm e} n_{\rm e} \D{t}{\bb{u}_{\rm e}} = -\grad\bcdot\msb{P}_{\rm e} - en_{\rm e} \left(\bb{E}+\frac{\bb{u}_{\rm e}}{c}\btimes\bb{B}\right) ,
\end{equation}
after using the smallness of the electron mass $m_{\rm e}$ to neglect the inertial term on the left-hand side and specifying the form of the electron pressure tensor $\msb{P}_{\rm e}$. The electron number density $n_{\rm e}$ and flow velocity $\bb{u}_{\rm e}$ are determined by quasi-neutrality,
\begin{equation}\label{eqn:quasineutrality}
    e n_{\rm e} = \sum_i q_i n_i = \sum_i q_i \int\rmd\bb{v} \, f_i ,
\end{equation}
and the non-relativistic version of Amp\`{e}re's law,
\begin{align}\label{eqn:ampere}
    e n_{\rm e} \bb{u}_{\rm e} &= \sum_i q_i n_i\bb{u}_i - \bb{j} \nonumber\\*
    \mbox{} &= \sum_i q_i \int\rmd\bb{v}\, \bb{v}f_i  - \frac{c}{4\pi}\grad\btimes\bb{B} .
\end{align}
Assuming for simplicity that the electrons are isothermal and isotropic, so that the electron pressure tensor $\msb{P}_{\rm e}=n_{\rm e} T_{\rm e}\msb{I}$ with $T_{\rm e}= \const$, equations \eqref{eqn:emom}--\eqref{eqn:ampere} may be combined to obtain a generalized Ohm's law,
\begin{equation}\label{eqn:ohm}
    \bb{E} = - \frac{\sum_i q_i n_i \bb{u}_i}{\sum_i q_i n_i c }\btimes\bb{B}+ \frac{(\grad\btimes\bb{B}) \btimes\bb{B}}{4\pi\sum_i q_i n_i}  -T_{\rm e}\frac{\grad\sum_i q_i n_i}{\sum_i q_i n_i}  .
\end{equation}
Because Eq.~\eqref{eqn:ohm} implies flux freezing in the electron fluid, we add a hyper-resistive term, $-\eta_4 \nabla^4\bb{B}$, to the right-hand side of Eq.~\eqref{eqn:faraday} in order to dissipate magnetic energy at small scales.

Equations~\eqref{eqn:vlasov}--\eqref{eqn:ohm} are solved in the presence of a constant and uniform background magnetic field ${\bb{B}_0=B_0\ez}$ using the massively parallel, hybrid-kinetic, particle-in-cell code {\tt Pegasus++} \citep{Kunz2014,Arzamasskiy2023}. We have added to this code a module allowing for multiple populations of different ion species, treated either passively or actively depending on the user's needs. {\tt Pegasus++} features highly optimized algorithms, including a fully vectorized particle pusher utilizing a modified Boris scheme, an almost fully vectorized deposit function with second-order-accurate particle weighting, and constrained transport to update the magnetic field in a divergence-free manner. These are all adopted within a three-stage predictor-predictor-corrector time integration scheme. We employ units such that the proton skin depth $d_{\rm p0}$ and the proton Larmor frequency $\Omega_{\rm p0}$ in the background uniform state are set to unity; the Alfv\'en speed in the background plasma is then $v_{\rm A0}=B_0/\sqrt{4\pi m_{\rm p}n_{\rm p0}} = d_{\rm p0}\Omega_{\rm p0}=1$. The initial proton Larmor radius given in these units is $\rho_{\rm p0}=\beta_{\rm p0}^{1/2}$, where $\beta_{\rm p0} \doteq 8\pi n_{\rm p0} T_{\rm p0}/B^2_0$ and $T_{\rm p0}$ is the initial proton temperature.

The forcing fields $\bb{E}^U_{\rm ext}$ and $\grad\btimes\bb{E}^B_{\rm ext}$ are designed to inject energy and cross helicity into the system at rates $\varepsilon$ and $\varepsilon_H$, respectively. They consist of random combinations of large-scale Fourier modes with wavenumber $k_j$ satisfying $2\pi/L_j\leq k_j \leq 4\pi/L_j$ for $j=\{x,y,z\}$. These modes are divergence-free, perpendicular to $\bb{B}_0$ (``$\perp$''), and time-correlated via an Ornstein--Uhlenbeck process with correlation time $\tau_{\rm corr}$. By separately generating bulk velocity fluctuations $\bb{u}_\perp$ and magnetic-field fluctuations $\bb{B}_\perp$, the combinations $\bb{F}^{\pm} \doteq \bb{E}^{U}_{\rm ext} \pm \grad\btimes\bb{E}^{B}_{\rm ext}$ specify the injection of energy into the Els\"asser fields 
\begin{equation}
    \bb{z}^\pm \doteq \bb{u}_\perp \pm \frac{\bb{B}_\perp}{(4\pi m_{\rm p} n_{\rm p})^{1/2}} \doteq \bb{u}_\perp\pm\bb{b}_\perp ;
\end{equation}
using this convention, $\bb{z}^\pm$ perturbations propagate in the $\mp\ez$ direction. Hence, the ratio $\langle\bb{z}^{+}\bcdot\bb{F}^{+}\rangle/\langle\bb{z}^{-}\bcdot\bb{F}^{-}\rangle$ determines the level of imbalance injected at the largest scales, with $\varepsilon_{\rm inj}\doteq \frac{1}{2}\langle\bb{z}^+\bcdot\bb{F}^+\rangle +\frac{1}{2}\langle\bb{z}^-\bcdot\bb{F}^-\rangle$ being the total energy injection rate. Note that, because the driving is time-correlated rather than white noise, $\varepsilon_{\rm inj}$ varies in time and is not in general equal to $\varepsilon$.

This paper presents results from two simulations, both of which adopt $\beta_{\rm p0}=0.3$ and are conducted within a Cartesian, periodic, elongated box having $(280)^2\times 1680$ cells spanning a domain of physical size $(28\pi\rho_{\rm p0})^2\times 168\pi\rho_{\rm p0}$. The box shape approximately matches the (statistical) shape of turbulent eddies measured at similar scales in the solar wind \citep{chen2016r,Squire2022}. These parameters imply a maximum resolved wavenumber of $k\rho_{\rm p0}=10$, minimum parallel and perpendicular wavenumbers of $k_{\parallel,\rm min}\rho_{\rm p0}\approx 0.01$ and $k_{\perp,\rm min}\rho_{\rm p0}\approx 0.07$, and a dimensionless Alfv\'en box-crossing time of $\tau_{\rm A}\approx 289$. In both runs, there are $10^3$ macro-particles per cell representing the protons, and $64$ macro-particles per cell for each additional ion species. All ion species are initialized by sampling from isotropic Maxwellian VDFs having the same temperature, $T_i=T_{\rm p0}=T_{\rm e0}$, and distributing their macro-particles evenly throughout the domain. Both runs have $\varepsilon=12.9$ and $\tau_{\rm corr}=\tau_{\rm A}/4$, which result in critically balanced fluctuations at the outer scale with root-mean-square (rms) bulk velocity $u_{\rm rms}\approx 1/6$. The only difference between the two runs is that the ``balanced'' one has $\varepsilon_H=0$, while the ``imbalanced'' one aims for $\varepsilon_H = 0.9\varepsilon$ by setting $|\bb{F}^+|^2/|\bb{F}^-|^2 = (\varepsilon+\varepsilon_H)/(\varepsilon-\varepsilon_H)=19$.

For the ionic species, we focus primarily on protons (p), alphas ($\alpha$), and quintuply ionized oxygen atoms (O$^{5+}$). Alphas are chosen due to their being the second-most abundant species in the solar wind (with fractional abundances ${\sim}1\mbox{--}10\%$) and therefore relatively well diagnosed \citep[e.g.,][]{Kasper2007,marsch12,Verscharen19,Mostafavi22}. ${\rm O}^{5+}$ is included as a heavier ion for which remote sensing (UVCS) observations have shown extreme perpendicular heating compared to protons \citep{Kohl2006}; this species has also been included in some prior studies on stochastic heating \citep{Chandran2010a}. Other studies have looked at simulations with either monochromatic or a broadband spectrum of AWs/ICWs and decaying turbulence with ${\rm O}^{5+}$ \citep{Hellinger2005,Maneva2016} or alphas \citep{Maneva2015,Valentini2016,Shi2022}. The turbulent heating of additional ionic species relevant to the solar wind (e.g., C$^{5+}$, C$^{6+}$, O$^{6+}$, Fe$^{9+}$) will be analyzed in a separate publication.

To isolate the effects of imbalance and the helicity barrier on ion heating and enable a direct comparison with the previously published proton-only simulation \citep{Squire2022}, we treat the non-proton species passively, {\it viz.}, their abundances are taken to be so small that they do not contribute appreciably to the summations in Eqs.~\eqref{eqn:quasineutrality}--\eqref{eqn:ohm}. This is certainly a good approximation for ${\rm O}^{5+}$ in the solar wind, but may not be accurate for alphas, whose fractional abundances of up to ${\sim}10\%$ of the solar wind have been measured by PSP \citep{Mcmanus2022}. The self-consistency of our passive treatment of the non-proton species is discussed at the end of Section~\ref{sec:results}.

\section{Results}\label{sec:results}

\subsection{Helicity Barrier and Excitation of PCWs}

%%%%%%%%%%%%%%%%%%%%%%% FIGURE 1 %%%%%%%%%%%%%%%%%%%%%%%%%%
\begin{figure}
    \centering
    \includegraphics[width=0.95\columnwidth]{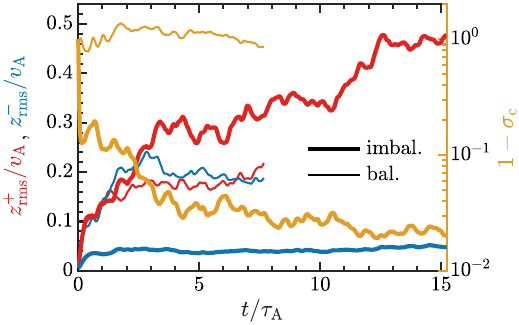}
    \caption{Time evolution of $z^{\pm}_{\rm rms}$ (red/blue) and $\sigma_{c}$ (yellow) throughout the imbalanced (thick lines) and balanced (thin lines) simulations.}
    \label{fig:sigma}
\end{figure}
%%%%%%%%%%%%%%%%%%%%%%%%%%%%%%%%%%%%%%%%%%%%%%%%%%%%%%%%%%

Because the non-proton species are treated passively, the development of imbalanced turbulence and the consequent helicity barrier in our imbalanced simulation is qualitatively similar to that presented in \citet{Squire2023}. The time evolution of the root-mean-square amplitudes $z^{\pm}_{\rm rms} \equiv \langle |\bb{z}^{\pm}|^{2} \rangle^{1/2}$ and the imbalance (normalized cross helicity),
\begin{equation}
    \sigma_{\rm c} \doteq \frac{\langle|\bb{z}^+|^2\rangle-\langle|\bb{z}^-|^2\rangle}{\langle|\bb{z}^+|^2\rangle+\langle|\bb{z}^-|^2\rangle} = \frac{2\langle\bb{u}_\perp\bcdot\bb{b}_\perp\rangle}{\langle|\bb{u}_\perp|^2\rangle+\langle|\bb{b}_\perp|^2\rangle},
\end{equation}
is shown in Fig.~\ref{fig:sigma}. As the large scales are forced, an imbalanced state with $\sigma_{\rm c} \approx 0.9$ and fluctuation amplitude predominantly in $z^{+}$ develops. A helicity barrier is set up within ${\approx} 2\tau_{\rm A}$, preventing the imbalanced portion of the energy in $z^{+}$ from cascading to scales smaller than $k_{\perp} \rho_{\rm p}  \sim 1$. Consequently, the energy in $z^{-}$ saturates very quickly, while the energy in $z^{+}$ and thus the imbalance continue to grow, saturating after $12 \tau_{\rm A}$ with $\sigma_{\rm c} \approx 0.98$. In the balanced run, whose evolution is traced by the thin lines in Fig.~\ref{fig:sigma}, the energy in $z^{+}$ and $z^{-}$ saturates within ${\approx}3\tau_{\rm A}$ and $\sigma_{\rm c} \approx \pm 0.1$ throughout. As a result, for a given energy injection rate the amplitude of turbulence in the saturated state is higher in the imbalanced run, $B_{\perp,\rm rms}/B_0 \approx 0.33$, than in the balanced run, $B_{\perp,\rm rms}/B_0\approx 0.19$.

%%%%%%%%%%%%%%%%%%%%%%% FIGURE  2 %%%%%%%%%%%%%%%%%%%%%%%%%%
\begin{figure}
    \centering
    \includegraphics[width=0.95\columnwidth]{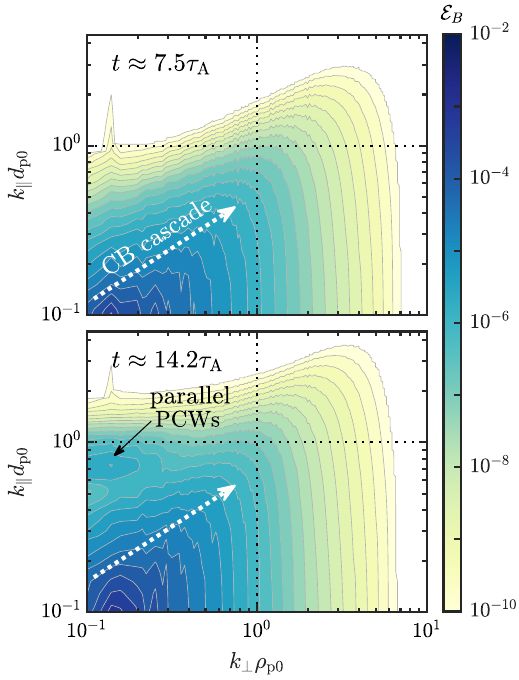}
    \caption{Spectra of perpendicular magnetic-field fluctuations, $\mathcal{E}_{B}= \mathcal{E}_{B_{y}} + \mathcal{E}_{B_{z}}$, in $(k_{\perp},k_{\parallel})$ space, with logarithmic color bar and contours, at intermediate and late times of the imbalanced simulation. The white dotted lines show the top edge of the critical-balance cone under which energy cascades to smaller scales; the black dotted lines mark $k_{\perp} \rho_{\rm p0}=1$ and $k_{\parallel} d_{\rm p0}=1$. The presence of parallel proton-cyclotron waves (PCWs) is indicated in the bottom panel.}
    \label{fig:spectra}
\end{figure}
%%%%%%%%%%%%%%%%%%%%%%%%%%%%%%%%%%%%%%%%%%%%%%%%%%%%%%%%%%

Fig.~\ref{fig:spectra} reveals a cascade of magnetic energy to smaller scales in the imbalanced run, occurring in $k_{\perp}$-$k_{\parallel}$ space predominantly within a critically balanced cone demarcated by $k_\parallel\propto z^+_{\rm rms}k^{2/3}_\perp$ (white dotted line).\footnote{These 2D spectra are calculated using the field-line-following method outlined in \citet{Squire2022}.} The helicity barrier causes $z^+_{\rm rms}$ to increase due to throttling of the imbalanced flux above $k_{\perp} \rho_{\rm p} \approx 1$, and the cone shifts upwards in $k_{\parallel}$ with time. This upwards shift means that a significant proportion of turbulent energy flows towards $k_{\parallel} d_{\rm p0} \sim 1$ before reaching $k_{\perp} \rho_{\rm p0} \sim 1$, causing these Alfv\'enic fluctuations to acquire the characteristics of oblique proton-cyclotron waves (PCWs). Diffusion of protons along the resonance contours of these oblique PCWs then occurs \citep{KennelEngelmann1966}, allowing the turbulence to saturate as the protons absorb the remaining energy input \citep{Li99}. The proton VDF ultimately becomes sufficiently unstable to generate power in parallel-propagating PCWs \citep{KennelWong1967,Chandran2010b}, which are apparent at small $k_{\perp}\rho_{\rm p0}$ and $k_\parallel d_{\rm p0} \approx 0.6\mbox{--}1$ in the bottom panel of Fig.~\ref{fig:spectra} ($t\approx 14.2\tau_{\rm A}$).

%%%%%%%%%%%%%%%%%%%%%%% FIGURE 3 %%%%%%%%%%%%%%%%%%%%%%%%%%
\begin{figure}
    \centering
    \includegraphics[width=0.95\columnwidth]{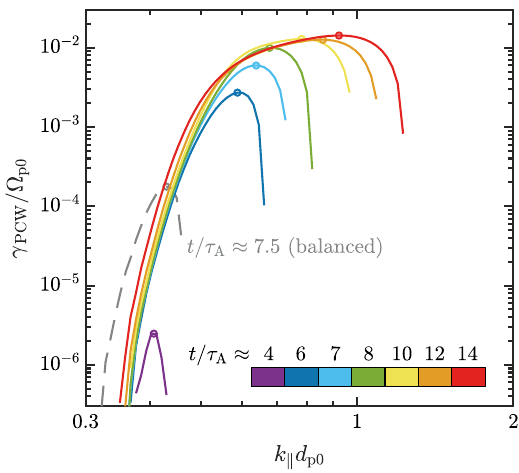}
    \caption{Linear growth rates of parallel PCWs, $\gamma_{\rm \,PCW}$, versus parallel wavenumber, $k_\parallel d_{\rm p0}$, as determined by \texttt{ALPS}, given the gyro- and box-averaged proton VDFs measured in the simulations at the stated times. The colored solid lines refer to the imbalanced run; the grey dashed line refers to the end of the balanced run.}
    \label{fig:PCWgrowth}
\end{figure}
%%%%%%%%%%%%%%%%%%%%%%%%%%%%%%%%%%%%%%%%%%%%%%%%%%%%%%%%%%

To elucidate further the presence of these parallel PCWs, we calculate their linear growth rates using the Arbitrary Linear Plasma Solver (\texttt{ALPS}) \citep{Verscharen2018,alps2023}. \texttt{ALPS} solves the hot-plasma kinetic dispersion relation and searches for unstable modes, with gyro- and box-averaged proton VDFs taken from the two simulations used as input (examples of these VDFs are provided and discussed later in this section; see Fig.~\ref{fig:vdfs} and accompanying text). The colored lines in Fig.~\ref{fig:PCWgrowth} trace the growth rates $\gamma_{\rm PCW}$ vs.~$k_\parallel$ of the parallel PCWs that are rendered linearly unstable by the proton VDFs occurring at different times in the imbalanced turbulence. Because wave-particle interactions in the imbalanced turbulence cause these VDFs to become asymmetric, the PCWs propagating in the direction of $z^-$ (the smaller-amplitude Els\"asser field) are either stable or much more slowly growing than those propagating in the direction of $z^+$, and the so the growth rates displayed in the figure pertain to the latter. Although this mode is unstable at early times in the imbalanced run, its growth rate is not large enough to compete with the turbulence (the Alfv\'en frequency $k_\parallel v_{\rm A0}$ at the start of the inertial range is ${\sim}10^{-2}\Omega_{\rm p0}$). But by time $t\gtrsim 10\tau_{\rm A}$, the growth rate of these parallel PCWs is large enough to produce significant power in the wavenumber range $k_\parallel d_{\rm p0} \approx 0.6\mbox{--}1$, consistent with the bottom panel of Fig.~\ref{fig:spectra}. In contrast, the proton VDFs in the balanced run are never distorted enough to yield appreciable parallel PCW growth rates (see the grey dashed line in Fig.~\ref{fig:PCWgrowth}, which corresponds to the end of the balanced run).

\begin{figure*}
    \centering
    \includegraphics[width=0.95\textwidth]{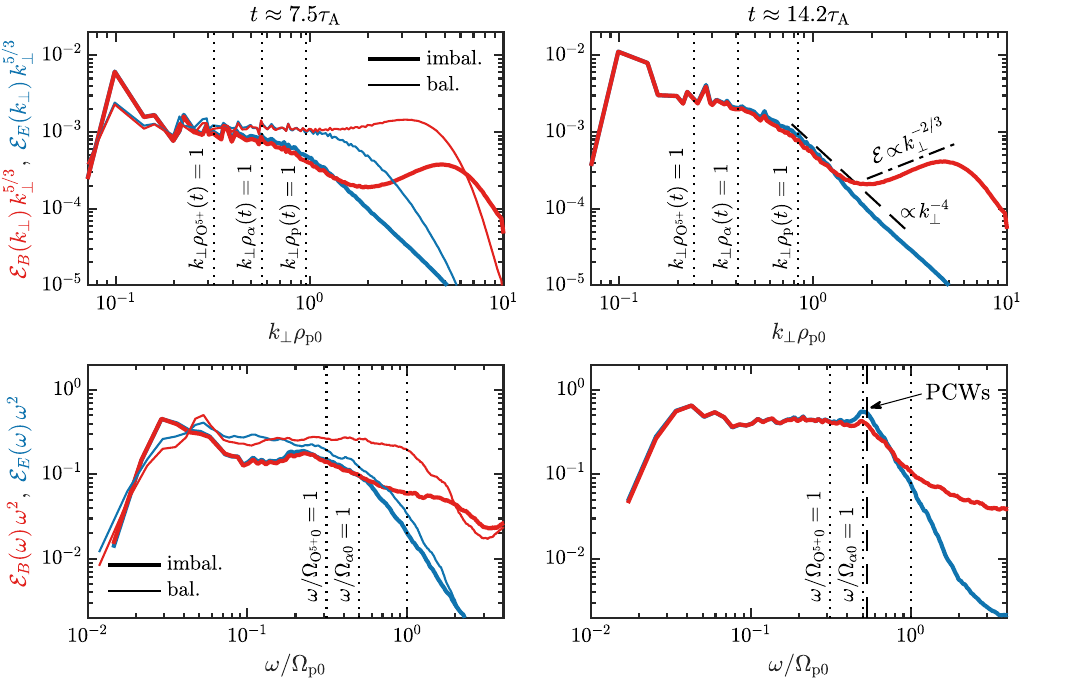}
    \caption{1D spectra of perpendicular magnetic-field and electric-field fluctuations, $\mathcal{E}_{B}$ (blue) and $\mathcal{E}_{E}$ (red), plotted against $k_{\perp}$ (top) and frequency $\omega$ (bottom) at intermediate (left) and late (right) times, in the imbalanced simulation (thick lines). The steady-state spectra from the balanced simulation are included in the left panels for comparison (thin lines). Wavenumber (frequency) spectra are compensated by $k^{5/3}_\perp$ ($\omega^2$). Wavenumbers and frequencies corresponding to the ion species' inverse-Larmor scales and Larmor frequencies at each time are indicated by the dotted lines. The vertical dashed line in the bottom-right panel labelled ``PCWs'' marks the real frequency of unstable PCWs calculated by \texttt{ALPS} at $t\approx 14.2\tau_{\rm A}$.}
    \label{fig:spectra1d}
\end{figure*}

The helicity barrier and the excitation of high-frequency fluctuations can also be identified in the 1D energy spectra at the two times shown in Fig.~\ref{fig:spectra1d}. Note that the wavenumber spectra are compensated by $k^{5/3}_\perp$ and the frequency spectra are compensated by $\omega^2$, so that a horizontal line on these plots corresponds to predictions for a constant-flux, critically balanced cascade of Alfv\'enic fluctuations \citep{Goldreich1995}. In the top panels, the presence of a helicity barrier in the imbalanced run (thick lines) gives rise to a characteristic steepening of the $k_\perp$-space spectra just above $k_{\perp} \rho_{\rm p0} \sim 1$, a ``transition range'' with $\mathcal{E}\propto k^{-4}_\perp$, and a re-flattening of the electric spectrum at sub-$\rho_{\rm p0}$ scales to $\mathcal{E}\propto k^{-2/3}_\perp$. These features are consistent with near-Sun PSP observations \citep{Bowen2020b,McIntyre2024} showing a transition range in the magnetic spectrum, in which a steepening of the inertial-range scaling from $k^{-5/3}_\perp$ to $k^{-4}_\perp$ at ion-kinetic scales is followed at yet smaller scales by a spectrum consistent with $k^{-8/3}_\perp$. While our magnetic spectra do not recover a $k^{-8/3}_\perp$ scaling in the sub-ion-Larmor range due to resistive dissipation \citep[cf.][]{Squire2022}, some re-flattening does begin at $k_{\perp} \rho_{\rm p0} \sim 2$. These features are clearly distinguishable when compared against the $k_\perp$-space spectrum from the balanced run (thin lines), whose inertial range follows $\mathcal{E}\propto k^{-5/3}_\perp$ all the way until $k_\perp\rho_{\rm p} \approx 1$, beyond which the fluctuations mutate into a cascade of dispersive kinetic Alfv\'en waves \citep[cf.][]{Howes08b,Arzamasskiy2019,Cerri2021}.

%%%%%%%%%%%%%%%%%%%%%%% FIGURE 4 %%%%%%%%%%%%%%%%%%%%%%%%%%
%%%%%%%%%%%%%%%%%%%%%%%%%%%%%%%%%%%%%%%%%%%%%%%%%%%%%%%%%%

The corresponding (compensated) frequency spectra are shown in the bottom row of Fig.~\ref{fig:spectra1d}. That the low-frequency ends of the balanced-run spectra approximately satisfy $\mathcal{E}\propto\omega^{-2}$ is consistent with a near-constant flux of turbulent energy to smaller scales (larger frequencies) \citep{Corrsin1963,Beresnyak2015,Schekochihin2022}. The imbalanced-run spectra at $t\approx 7.5\tau_{\rm A}$ are clearly different, consistent with the idea that conservation of generalized helicity precludes a constant-flux cascade for the imbalanced portion of the turbulence (this must be the case given that $z^+_{\rm rms}$ is still increasing at this time). At the later time, $t\approx 14.2\tau_{\rm A}$, the barrier saturates and a constant-flux spectrum emerges at frequencies  $\omega\lesssim 0.4\Omega_{\rm p0}$. A high-frequency bump appears about $\omega/\Omega_{\rm p0}\approx 0.52$, a number that matches the real frequency of PCWs as calculated by \texttt{ALPS} at $t\approx 14.2\tau_{\rm A}$ (marked by the vertical dashed line in the figure) Building upon previous studies, we have elaborated on signatures of the helicity barrier observed in prior {\tt Pegasus++} simulations \citep{Squire2022,Squire2023}, and gleaned additional insight via linear calculations. In particular, we have determined the linear plasma response using the non-bi-Maxwellian proton VDFs taken from the simulations, and found excellent agreement with the properties of outwardly propagating, unstable, parallel PCWs in this strongly turbulent non-linear system.

\subsection{Phase-space Signatures of Ion Heating}

\begin{figure*}
    \centering
    \includegraphics[width=0.956\textwidth]{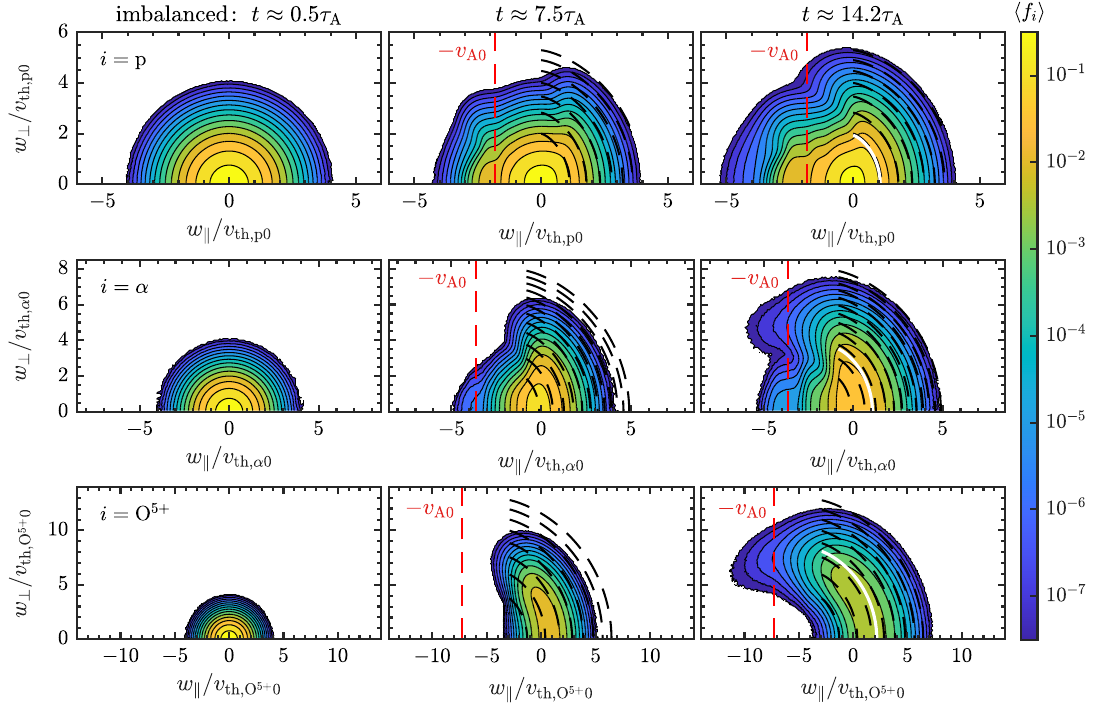}
    \caption{VDFs, $f_i(w_\parallel,w_\perp)$, in peculiar velocity $(w_{\parallel}, w_{\perp})$ of protons (top row), alphas (middle row), and O$^{5+}$ (bottom row) at early (left column), intermediate (middle column), and late (right column) times in the imbalanced simulation, with logarithmic color bar and contours. The axes are scaled to the initial thermal speeds of each respective species, $\vth{,i0}$. The kinetic-Alfv\'en-wave Landau resonance is indicated by the red dashed line. The black dashed lines trace the oblique PCW resonance contours for each respective species; the white lines in the last column mark the resonance contour used in Fig.~\ref{fig:fprp-balanced}.}
    \label{fig:vdfs}
\end{figure*}

We now shift our attention to the main focus on this paper -- the differential heating of the ionic species. We begin by showing evidence for strong perpendicular ion heating in the presence of a helicity barrier. In \citet{Squire2022,Squire2023}, the mechanism responsible for perpendicularly heating the protons was postulated to be repeated instances of the proton VDF diffusing along oblique PCW resonances and relaxing through the parallel PCW instability, leading to diffusion across the oblique resonance contours \citep{Chandran2010b}. Evidence for this process can be seen in the evolution of the proton VDFs in the top panels of Fig.~\ref{fig:vdfs}. The initially Maxwellian proton VDF (top-left panel) diffuses along the oblique cyclotron contours (dashed lines\footnote{The resonant scattering contours are defined as the level sets of some function $\eta(w_\perp,w_\|)$ for which $\mathcal{G}_{\rm res}(\eta)=0$, where $\mathcal{G}_{\rm res}$ is the quasi-linear diffusion operator $\mathcal{G}_{\rm res}\equiv (1-w_{\|}/v_{\rm ph})\partial/\partial w_{\perp}+(w_{\perp}/v_{\rm ph})\partial/\partial w_{\|}$ \citep{KennelEngelmann1966}. Here, $v_{\rm ph}(k_\|,k_\perp)=\omega_{\rm PCW}/k_\|$ is the phase velocity of the oblique PCWs, which is itself a function of $w_\|$ via the resonance condition $w_\parallel = (\Omega_i - \omega_{\rm PCW})/|k_\parallel|$ for species $i$. We use the cold plasma, $k_\perp\gg k_\|$ approximation to the PCW dispersion relation, $\omega=k_{\|}v_{\rm A}/\sqrt{1+k_{\|}^{2}d_{\rm p}^{2}}$ \citep{Stix1992,Isenberg12}, to compute these contours for Figs~\ref{fig:vdfs} and~\ref{fig:fprp-balanced}. For protons, $\omega_{\rm PCW}<\Omega_{\rm p}$, implying that the resonant contours ``stop'' at $w_\|=0$ (where the resonant $k_\|\rightarrow\infty$); in contrast, $\omega_{\rm PCW}>\Omega_i$ for minor ions at high $k_\|$, with $w_\|=(\Omega_i - \omega_{\rm PCW})/|k_\parallel|$ reaching some negative minimum value at finite $|k_\||$ (then increasing with $|k_\||$ at larger $|k_\||$). This causes the resonant contours to extend past $w_\|=0$ for minor ions, as seen in Fig.~\ref{fig:vdfs}.}) in the top-middle panel at $t\approx 7.5\tau_{\rm A}$, before the protons have undergone significant heating. There is also a Landau resonance with the imbalanced Alfv\'enic fluctuations at $k_\perp\rho_{\rm p}\sim 1$, which flattens the VDF about $w_{\parallel} \approx -v_{\rm A}$ \citep{Howes2006}, forming a modestly super-Alfv\'enic beam feature in the direction of dominant wave propagation similar to those observed in the fast solar wind \citep{Marsch1982a,Marsch2006,Verniero20}. At later times (top-right panel), the proton VDF spreads across the oblique PCW contours as parallel PCWs are excited, a process that cools the plasma by transferring some of the available free energy in the unstable VDF into electromagnetic fluctuations. However, because the parallel PCW resonance contours are more curved relative to the oblique ones, the generation of parallel PCWs and subsequent diffusion along their resonance contours, can bring protons onto another oblique contour resonance at yet higher energy \citep{Chandran2010b,Squire2022}. This leads to further perpendicular heating by the oblique PCWs. Through this transfer of electromagnetic energy to the thermal energy of the protons, the perpendicular proton temperature increases to $T_{\perp, \rm p} \approx 1.6T_{\perp, \rm p0}$ and the electromagnetic energy in the imbalanced turbulence saturates.
%
%%%%%%%%%%%%%%%%%%%%%%% FIGURE 5 %%%%%%%%%%%%%%%%%%%%%%%%%%
%%%%%%%%%%%%%%%%%%%%%%%%%%%%%%%%%%%%%%%%%%%%%%%%%%%%%%%%%%

The lower two rows of panels in Fig.~\ref{fig:vdfs} illustrate a similar process -- albeit more extreme -- for the other ion species. Note that the axes are scaled according to each species' initial thermal speed, $\vth{,i0}\doteq (2T_{i0}/m_i)^{1/2} = \vth{,\rm p0}(m_p/m_i)^{1/2}$. By the end of the run, the alphas and O$^{5+}$ ions are perpendicularly heated substantially more than the protons, with $T_{\perp,\alpha}/T_{\alpha 0}\approx 6.4$ and $T_{\perp,{\rm O}^{5+}}/T_{{\rm O}^{5+}0}\approx 28$ giving $T_{\perp, \alpha}/T_{\perp, \rm p} \approx 4.1$ and $T_{\perp, \rm O^{5+}}/T_{\perp, \rm p} \approx 18$. Just as for the protons, these other VDFs diffuse along curves corresponding to their resonance conditions with oblique Alfv\'en fluctuations having $k_{\parallel} v_{\rm A} \sim \Omega_{i}$. Because these fluctuations cyclotron damp similarly to the oblique PCWs at $k_{\parallel} d_{\rm p}\sim 1$, we also refer to them as oblique PCWs. The resonance contours for ions interacting with these oblique PCWs (dashed lines) are steeper in $w_\perp/\vth{i0}$ for the heavier species and therefore allow them to experience more perpendicular heating. This feature can be seen especially in the bottom-middle panel at $t\approx 7.5 \tau_{\rm A}$, in which the $\rm O^{5+}$ VDF is stretched vertically in $w_{\perp}$ relative to its initial Maxwellian state on the bottom-left by a substantial amount. The parallel PCWs play a less important role in the heating of alphas and minor ions, given the smaller $\Omega_{i}$ of these species; indeed, the strong heating of the non-proton species begins well before parallel PCWs are present.

Another interesting feature of the oblique PCW resonance curves is that they can exist to the left of the origin, $w_{\parallel} < 0$, for the heavier species. This is because their smaller charge-to-mass ratios can place their Larmor frequencies below the real frequency of the PCWs, causing their resonances with the dominant $z^+$ fluctuations to occur at parallel velocities $w_\parallel = (\Omega_i - \omega_{\rm PCW})/|k_\parallel|$ that are negative. Consequently, because minor ions with small $|w_{\parallel}|$ can lie within these curves, a greater population of minor ions (relative to protons) can be resonant. The cyclotron-resonant heating is in fact strong enough to push the minor-ion VDFs all the way over to the Landau resonance at $w_\parallel \approx -v_{\rm A}$. The consequent parallel heating then further accentuates the already arced shape of the VDFs, a feature that is most evident in the final (bottom-right) panel of Fig.~\ref{fig:vdfs} showing the final VDF of O$^{5+}$.

%%%%%%%%%%%%%%%%%%%%%%% FIGURE 6 %%%%%%%%%%%%%%%%%%%%%%%%%%
\begin{figure}
    \centering
    \includegraphics[width=0.95\columnwidth]{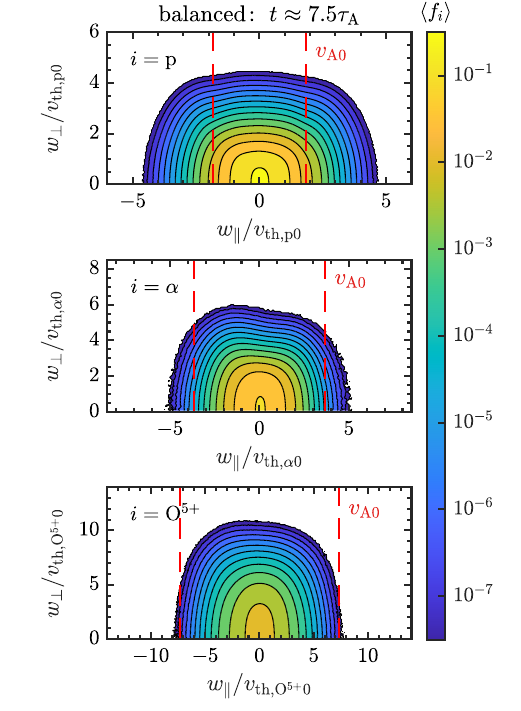}
    \caption{VDFs $f_i(w_\parallel,w_\perp)$ for $i={\rm p}$, $\alpha$, ${\rm O}^{5+}$ at a late time in the balanced simulation, with logarithmic color bar and contours. The axes are scaled to the initial thermal speed of each respective species, $v_{{\rm th}i0}$. The VDFs are approximately symmetric across $w_\parallel=0$ and flattened along $w_\perp$ in a manner indicative of stochastic heating (cf.~Fig.~\ref{fig:vdfs}). The kinetic-Alfv\'en-wave Landau resonances are indicated by the red dashed lines.}
    \label{fig:vdfs-balanced}
\end{figure}
%%%%%%%%%%%%%%%%%%%%%%%%%%%%%%%%%%%%%%%%%%%%%%%%%%%%%%%%%%

In contrast, the ion VDFs from the run with balanced driving, displayed in Fig.~\ref{fig:vdfs-balanced}, show no evidence of strong cyclotron-resonant heating. Because no helicity barrier is formed in this run, the inertial-range fluctuations remain at small enough amplitudes that the characteristic nonlinear frequency at scale $k_\perp$, {\em viz.}~${\sim}k_\perp z^{\pm}_{{\rm rms},k_\perp}$, is small compared to ${\sim}\Omega_{\rm p0}$. With critical balance implying comparable linear and nonlinear frequencies, this implies negligible power in PCWs, as observed. Instead, the ions appear to be mostly heated stochastically off of finite-amplitude electrostatic fluctuations at $k_\perp\rho_i\sim 1$ \citep{Chandran2010a}, although the total ion heating rate is significantly less than in the imbalanced run, with most (${\approx}60\%$) of the injected energy dissipating on hyper-resistivity (see below). The clearest evidence for this heating channel is the characteristic flat-top distributions seen in $w_{\perp}$ \citep[cf.][]{KleinChandran2016,Cerri2021}. The protons additionally experience appreciable parallel heating via Landau resonance  with the $k_\perp\rho_{\rm p0}\sim 1$ kinetic Alfv\'en waves, which flattens their VDF about $w_\parallel/v_{\rm th,p0}\approx \pm v_{\rm A0}$. At $\beta_{\rm p0}=0.3$, this resonance lies in the tail of the non-proton VDFs, so the parallel heating of these species is not as strong. The salient features of this proton VDF were also found in a previous {\tt Pegasus} simulation of balanced turbulence at $\beta_{\rm p}=0.3$ and analyzed in detail \citep[][see their figures~5--8]{Arzamasskiy2019}.

Fig.~\ref{fig:fprp-balanced} makes these features more evident and quantitative by providing various 1D VDFs. Its top panel shows the proton, alpha, and O$^{5+}$ VDFs at the end of the imbalanced run evaluated along the oblique-PCW resonant contour that starts at $(w_\parallel,w_\perp)=(v_{{\rm th}\parallel,i},0)$ for each species and arcs upwards towards smaller $w_\parallel$, where $v_{{\rm th}\parallel,i}\doteq(2T_{\parallel,i}/m_i)^{1/2}$ is the parallel thermal speed at $t\approx 14.2\tau_{\rm A}$; these contours are marked in the right-most panels of Fig.~\ref{fig:vdfs} by the white lines. The arc length along this contour is denoted by $s$, and is normalized by the instantaneous perpendicular thermal speed, $v_{{\rm th}\perp,i}\doteq (2T_{\perp,i}/m_i)^{1/2}$, for each species at $t\approx 14.2\tau_{\rm A}$. All species' VDFs are remarkably flat along the resonant contour. The bottom two panels show the perpendicular VDFs, $f(w_\perp)\equiv\int\rmd w_\parallel\, f(w_\parallel,w_\perp)$, corresponding to the 2D VDFs shown in Figs~\ref{fig:vdfs} and~\ref{fig:vdfs-balanced}; the initial Maxwellian is denoted by the dotted line, and the perpendicular velocities and VDFs are normalized using the instantaneous value of $v_{{\rm th}\perp,i}(t)$ at the times given. Though all of the 2D VDFs in the imbalanced run flatten along the oblique-PCW contours as they arc upwards in $w_\perp$ and leftwards in $w_\parallel$, the heavier ions are able to diffuse farther into $w_\parallel < 0$ because of their smaller charge-to-mass ratio. As a result, the corresponding perpendicular distributions are not universal, with $\langle f_{\rm p}(w_\perp)\rangle$ appearing to be nearly Maxwellian (despite the interesting structure exhibited in Fig.~\ref{fig:vdfs}) and $\langle f_{\rm O^{5+}}(w_\perp)\rangle$ exhibiting a local maximum near $w_\perp\approx v_{\rm th\perp,O^{5+}}$ that results from cyclotron-resonant diffusion along PCW contours that curve (leftwards in the figure) into the parallel direction. In contrast, all of the species in the balanced run exhibit almost identical $\langle f_i(w_\perp)\rangle$, with a clear flattening evident at sub-thermal perpendicular velocities that is strongly indicative of stochastic heating.

%%%%%%%%%%%%%%%%%%%%%%% FIGURE 7 %%%%%%%%%%%%%%%%%%%%%%%%%%
\begin{figure}
    \centering
    \includegraphics[width=0.95\columnwidth]{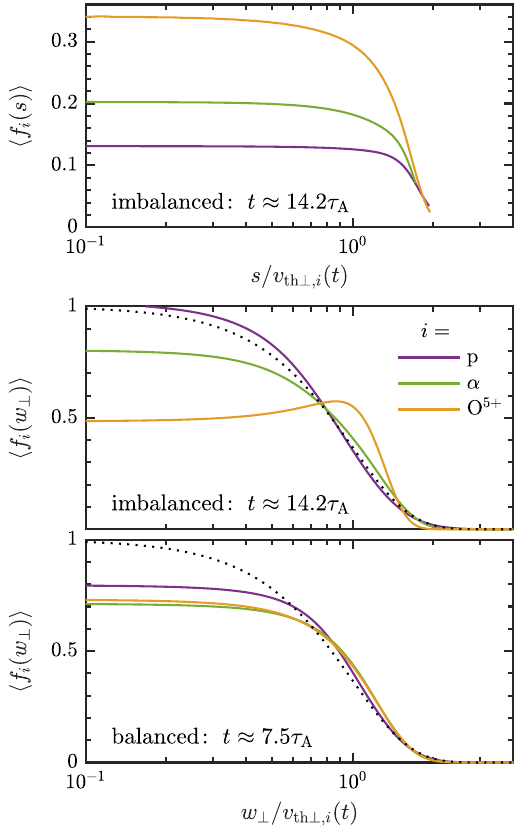}
    \caption{Top: $f_i(s)$ for $i={\rm p}$ (purple), $\alpha$ (green), ${\rm O}^{5+}$ (orange) at a late time in the imbalanced simulation, where $s$ is the arc length along the oblique-PCW resonant contour indicated by the white line for each species in the right-most panels of Fig.~\ref{fig:vdfs}. Bottom: $f_i(w_\perp)$ at late times in the imbalanced and balanced simulations; the black dotted line shows the initial Maxwellian. In all panels, the abscissas are scaled to the time-dependent perpendicular thermal velocity of each respective species and the VDFs are re-normalized accordingly.}
    \label{fig:fprp-balanced}
\end{figure}
%%%%%%%%%%%%%%%%%%%%%%%%%%%%%%%%%%%%%%%%%%%%%%%%%%%%%%%%%%

%%%%%%%%%%%%%%%%%%%%%%% FIGURE 8 %%%%%%%%%%%%%%%%%%%%%%%%%%
\begin{figure*}
    \centering
    \includegraphics[width=0.95\textwidth]{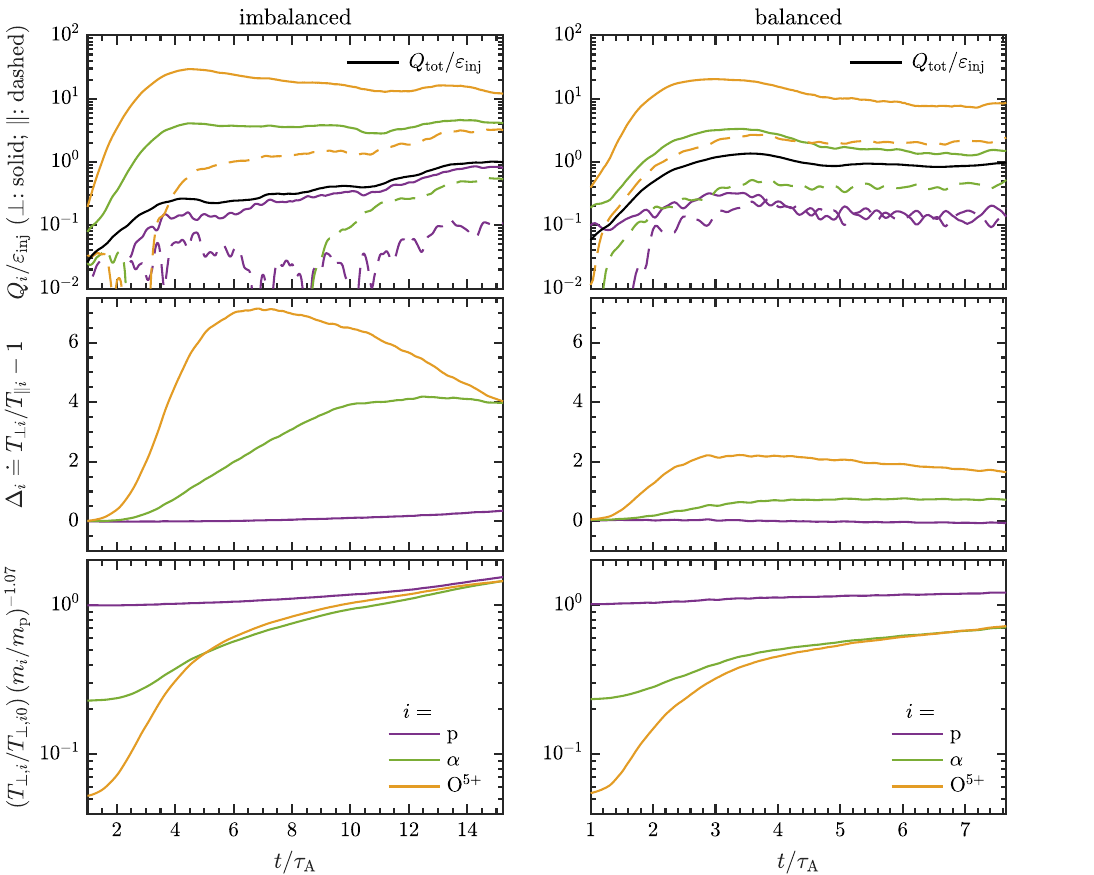}
    \caption{Top: Time evolution of heating rates, perpendicular $Q_{\perp,i}$ (solid) and parallel $Q_{\parallel,i}$ (dashed), for protons (purple), alphas (green), and O$^{5+}$ (orange) in the imbalanced (left) and balanced (right) simulations. The black lines trace the total plasma heating rate, including resistive dissipation but excluding the passive ions. Middle: Time evolution of temperature anisotropy $\Delta_i \doteq T_{\perp i}/T_{\parallel i}-1$ of each species. Bottom: Time evolution of perpendicular temperatures $T_{\perp,i}$, normalized using an empirical power-law scaling obtained from solar-wind measurements taken by ACE/SWICS at 1~au of collisionally young minor ions \citep{Tracy2016}.
    }
    \label{fig:heatingrates}
\end{figure*}
%%%%%%%%%%%%%%%%%%%%%%%%%%%%%%%%%%%%%%%%%%%%%%%%%%%%%%%%%%

\subsection{Differential Heating Rates of Ions}

The perpendicular and parallel heating rates and temperatures of each ion species in the imbalanced and balanced simulations are shown by the colored lines in Fig.~\ref{fig:heatingrates}, the former calculated via
\begin{gather}
    Q_{\perp i} \doteq \D{t}{}\int\rmd^3\bb{r}\int\rmd^3\bb{w}\, \frac{1}{2} m_i w^2_\perp f_i , \label{eqn:Qprp}\\*
    Q_{\parallel i}\doteq \D{t}{}\int\rmd^3\bb{r}\int\rmd^3\bb{w} \, \frac{1}{2} m_i w^2_\parallel f_i. \label{eqn:Qprl}
\end{gather}
These rates are normalized using the energy injection rate at the large scales, $\varepsilon_{\rm inj}$, averaged over the turbulent steady state of each run. The sum of the proton heating rates and the hyper-resistive dissipation rate, $Q_{\rm tot}$, is denoted by the black line; it omits the heating rates of the passive ion species. With the abundances of the passive species artificially set to unity in Eqns~\eqref{eqn:Qprp} and~\eqref{eqn:Qprl}, their heating rates in Fig.~\ref{fig:heatingrates} serve as an analogue for how rapidly their temperatures increase relative to the protons. 

In both runs, the heavier ion species are preferentially heated, especially so in the imbalanced case where the heavier species start to heat perpendicularly earlier in time than the protons. This is consistent with our analysis of the oblique-PCW resonance contours, with which a larger proportion of the non-proton VDFs can be in resonance at small $|w_\parallel|$. In addition, because heavier ions are in resonance with fluctuations at relatively longer wavelengths (e.g., $k_{\parallel} v_{\rm A}\sim \Omega_{\rm O^{5+}}$) where their amplitudes are larger, their heating rates are larger as well. 
The perpendicular heating rates of the alphas and O$^{5+}$ ions climb sharply as soon as sufficiently high frequencies (${\sim}\Omega_i$) are realized in the imbalanced cascade, because the resonance contours at these early times are significantly steeper than the isocontours of the still-nearly-Maxwellian VDFs. These rates peak after the onset of the helicity barrier when the increase in inertial-range fluctuation amplitudes stalls temporarily (at $t\approx 4\tau_{\rm A}$; see Fig.~\ref{fig:sigma}). Although these fluctuations then continue to increase in amplitude, and therefore parallel wavenumber, the alpha and minor-ion VDFs have already begun flattening along the oblique-PCW resonance contours. Quasi-linear theory suggests that the heating should then subside. 
Interestingly, however, the heating rates of the alphas and O$^{5+}$ ions either decrease only slightly in time or remain steady. Inspection of their VDFs in Fig.~\ref{fig:vdfs} indicates this continued heating is associated with particle diffusion across the oblique-PCW resonance contours. This diffusion occurs because the increased amplitudes and frequencies above the helicity barrier, which push the cone of critically balanced fluctuations up in $k_\parallel$ (Fig.~\ref{fig:spectra}), imply significant inertial-range power at the slower cyclotron frequencies of the heavier species at low $k_{\perp}$. The associated near-parallel cyclotron-resonance contours are flatter in $w_\perp$ than the oblique-PCW resonance contours. Consequently, the heavier species are able to leap from one oblique contour to the next so that heating persists.

Because of their smaller mass, only protons having $|w_\parallel|>0$ can be cyclotron-resonant with oblique PCWs. As a result of this relatively smaller population, the relatively shallower resonance contours, and the smaller fluctuation amplitudes with which the protons are resonant, the perpendicular heating of protons is more moderate and gradual throughout the simulation. There is a marked increase at later times as the proton VDF becomes increasingly unstable to generating parallel PCWs, and resonant interactions with these unstable parallel PCWs cause diffusion across the oblique-PCW contours. This heating is then sustained by repeated diffusion across oblique-PCW resonance contours via the parallel-PCW resonances.
The Landau damping of the proton core at $w_\parallel\approx -v_{\rm A0}$ leads to some initial proton parallel heating, which decreases after $6 \tau_{\rm A}$ when the VDF flattens across this resonance. For the non-proton species, initial parallel heating via the Landau resonance is negligible due to their initially sub-Alfv\'enic thermal speeds. Instead, appreciable parallel heating occurs, from ${\gtrsim}4\tau_{\rm A}$ for O$^{5+}$ and ${\gtrsim}9\tau_{\rm A}$ for alphas, once a non-negligible fraction of their broadened VDFs sample parts of the oblique PCW resonance contours at higher $w_{\perp}$ that curve towards the $-w_\parallel$ direction, leading to the arced VDFs. A secondary increase in parallel heating then occurs for all species, from ${\gtrsim}11\tau_{\rm A}$ for O$^{5+}$ and ${\gtrsim}12\tau_{\rm A}$ for protons and alphas, once the hoods of their arced VDFs approach and cross the Landau resonance.

In the balanced simulation, the heating rates reach a peak at $t\approx 3 \tau_{\rm A}$, once the turbulence has fully developed, and slowly decrease thereafter as the ion VDFs broaden perpendicularly from their initial Maxwellians. As a result of the heating, the temperatures of the minor ions in the imbalanced run exhibit a strong perpendicular bias, with the temperature anisotropy $\Delta_i\doteq T_{\perp,i}/T_{\parallel,i}-1$ peaking at ${\approx}7$ for $i={\rm O}^{5+}$ and ${\approx}4$ for $i=\alpha$ (Fig.~\ref{fig:heatingrates}, middle row). In the balanced run, the proton temperature is close to isotropic from the additional contribution of the parallel Landau-resonant heating, while the non-proton species attain more modest values of $\Delta_\alpha \approx 0.7$ and $\Delta_{{\rm O}^{5+}}\approx 2$. It is important to emphasize that the beam and core components of the species are not separated in these plots; for example, the heating of the proton core is almost entirely perpendicular, with the parallel heating being largely associated with the parallel beam at $w_\parallel \approx -v_{\rm A0}$ that forms due to Landau resonance with $k_\perp\rho_{\rm p}\sim 1$ kinetic Alfv\'en waves. Deciphering particle energization based solely on the bulk temperatures and temperature anisotropies, rather than by examining the structure of the VDF to separate the core and beam components, may confuse the distinct physical processes responsible for particle heating, a point worth bearing in mind when analyzing solar-wind data.

Finally, the bottom row of Fig.~\ref{fig:heatingrates} shows the increase in the perpendicular temperature of the species, normalized using an empirical power-law fit by \citet{Tracy2016} to measurements of the collisionally young solar wind taken by ACE/SWICS at 1~au, {\em viz.} $T_{i}/T_{\rm p} \propto (m_{i}/m_{\rm p})^{1.07}$. This normalization is remarkably successful at describing the late-time temperature evolution in the imbalanced run (the linear fit of $T_{i}/T_{\rm p}=1.35 m_{i}/m_{\rm p}$ also given by \citet{Tracy2016} gives slightly poorer agreement). This normalization also does well for describing the temperatures of the non-proton species in the balanced run, but fails for the protons (at least for the duration of this simulation; agreement might improve at yet larger times). The better fit for the imbalanced run lends further credence to the idea that minor ions might serve as a discriminating probe of turbulence and kinetic physics in the solar wind \citep{Bochsler2007}.

The partitioning of the turbulent energy into electron and proton heat was the focus of recent work on the helicity barrier and its gradual dissolution with temporally decreasing imbalance \citep{Squire2023}. Here we briefly make contact with that work.  Using the rate of hyper-resistive dissipation of the small-scale ($k_\perp\rho_{\rm p}\gg 1$) magnetic energy $\varepsilon_\eta$ as a proxy for the electron heating rate $Q_{\rm e}$, \citet{Squire2023} found quantitative agreement with the theoretical expectation that $Q_{\rm e} \approx \varepsilon - \varepsilon_H$ when the helicity barrier is active, i.e., that the balanced portion of the energy flux that is able to pass through the barrier cascades to sub-ion-Larmor scales and heats the electrons. Averaging over the steady state of our imbalanced run ($t/\tau_{\rm A}\gtrsim 12.5$), we find by examining the difference between the black line ($Q_{\rm tot}/\varepsilon_{\rm inj}$) and the sum of the purple lines ($Q_{\rm p}/\varepsilon_{\rm inj}$) in the top-left panel of Fig.~\ref{fig:heatingrates} that $Q_{\rm e}/\varepsilon_{\rm inj}\approx 0.07$, which matches well the theoretical expectation using the measured value of $1-\varepsilon_H/\varepsilon\approx 0.08$. The corresponding proton-to-electron heating ratio of ${\approx}14$ is much larger than in the balanced run, for which $Q_{\rm p}/Q_{\rm e}\approx 0.7$, a value consistent with $Q_{\rm p}/Q_{\rm e}$ measured at the end of the run in \citet{Squire2023} when $\varepsilon_H/\varepsilon\approx 0$ and the barrier had dissolved.\footnote{The $\beta_{\rm p0}=0.3$ balanced simulation in \citet{Arzamasskiy2019} returned a larger value of $Q_{\rm p}/Q_{\rm e}\approx 3$. Those authors attributed their ion heating to a combination of Landau and transit-time damping, stochastic heating off of $k_\perp\rho_{\rm p}\sim 1$ fluctuations, and, predominantly, cyclotron heating off of kinetic-Alfv\'en-wave fluctuations on scales $k_\perp\rho_{\rm p}\approx 4\mbox{--}5$ whose spectral anisotropies implied significant power at frequencies comparable to $\Omega_{\rm p0}$. The latter mechanism is not as dominant in our balanced simulation, presumably due to a combination of (i) an increased inertial range afforded by our slightly larger box size, which serves to decrease the amplitudes and increase the wavevector anisotropies of the fluctuations reaching sub-ion-Larmor scales; and (ii) our slightly larger hyper-resistivity, which cuts off the kinetic-Alfv\'en-wave cascade starting at $k_\perp\rho_{\rm p0}\approx 3$ rather than at the value $k_\perp\rho_{\rm p0}\approx 5$ realized in \citet{Arzamasskiy2019}.} Together, these results suggest that extreme heating of minor ions, enhanced proton-to-electron temperature ratios, outwardly propagating PCWs, and asymmetrically arced ion VDFs should correlate positively with the degree of imbalance.

\subsection{Potential Impact of Active Alphas and Minor Ions}

Because we treat the non-proton ions as passive, one might be concerned that the significant deviations from local thermodynamic equilibrium seen in the $\alpha$ and O$^{5+}$ VDFs could serve as a free-energy source to inject electromagnetic energy back into the plasma and/or modify the properties of the excited PCWs. To investigate this possibility, we again employ \texttt{ALPS}, this time using as input the gyro- and box-averaged VDFs from all ion species taken at several different times in the two simulations. For each time, we calculate the complex frequencies $\omega$ for both forward- and backward-propagating fast and Alfv\'en waves for a broad range of relative ion abundances. We find that no realistic abundance of O$^{5+}$ changes the stability of the plasma, so in what follows we focus exclusively on the impact of an active alpha population.

\begin{figure}
    \centering
    \includegraphics[width=0.95\columnwidth]{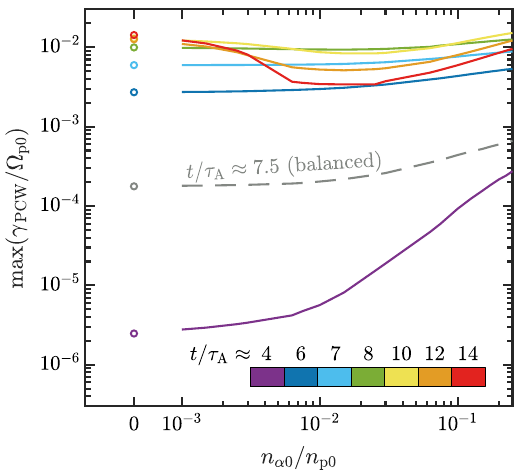}
    \caption{Maximum linear growth rate of parallel PCWs, $\gamma_{\rm \,PCW}$, versus alpha abundance as determined by \texttt{ALPS}, given the gyro- and box-averaged proton and alpha VDFs measured in the simulations at the stated times. The colored solid lines refer to the imbalanced run; the grey dashed line refers to the end of the balanced run.}
    \label{fig:PCWgrowth-alpha}
\end{figure}

The maximum growth rates of the linearly unstable PCWs as a function of active alpha abundance, $n_{\alpha 0}/n_{\rm p0}$, are provided in Fig.~\ref{fig:PCWgrowth-alpha}. The solid (dashed) lines pertain to the imbalanced (balanced) run, with the corresponding times indicated and marked by different colors. Because our simulations employ passive alphas, the open circles at $n_{\alpha 0}/n_{\rm p0}=0$ correspond to the maximum PCW growth rates shown in Fig.~\ref{fig:PCWgrowth}. For times $t\lesssim 8\tau_{\rm A}$, the impact of an alpha contribution is insignificant, only marginally changing the maximum growth rate and the region of wavevector support of the most unstable parallel PCW. (For the curve at $t/\tau_{\rm A}\approx 4$, the two orders of magnitude increase in $\gamma_{\rm PCW}$  with alpha abundance is still not enough to make these modes relevant.) At times $t\gtrsim 10\tau_{\rm A}$, when the parallel PCWs appear unambiguously in the fluctuation spectra of the imbalanced run, the alpha population impacts the unstable PCW modestly, with the growth rate of the most unstable mode reduced by a factor of ${\approx}2\mbox{--}3$ at the latest times for relative densities of ${\sim}$1\%--10\%. This reduction in growth rate is caused by the alphas resonantly absorbing a fraction of the power being emitted by the protons. Importantly, we find that at no time are alpha-cyclotron waves driven unstable, and that the oblique PCWs at $k_\parallel d_{\rm p0} \sim k_\perp \rho_{\rm p0} \sim 1$ down to $k_\parallel d_{\rm p0} \sim 0.3$ that are most responsible for heating the protons and minor ions have their damping rates (not shown in Fig.~\ref{fig:PCWgrowth-alpha}) increased by no more than a factor of ${\approx}3$ at the very largest alpha abundances.

This analysis supports the passive treatment of the alphas and minor ions in our simulation. Studying the nonlinear impact of a realistic population of alphas and minor ions for different plasma and turbulence parameters is an area for further research.

%%%%%%%%%%%%%%%%%%%%%%% FIGURE 9 %%%%%%%%%%%%%%%%%%%%%%%%%%

\section{Summary}
\label{sec:summary}

Because of their relatively low abundances and variety of masses and charge states, minor ions serve as excellent probes of the physics of waves, turbulence, and dissipation across different scales in the collisionless solar wind. This is especially true when knowledge of their temperatures and VDFs is combined with observations of fluctuation spectra, proton VDFs, and proton-to-electron heating ratios, affording a diverse suite of diagnostics capable of discriminating between different theories for how the solar wind is powered. Following \citet{Squire2022,Squire2023}, we have  demonstrated that the increased inertial-range fluctuation amplitudes that result from the helicity barrier in imbalanced turbulence can excite high-frequency fluctuations (namely, oblique and parallel PCWs) that efficiently heat solar-wind protons in a manner consistent with a wide range of solar-wind observations \citep[e.g., by PSP;][]{Bowen2020a,Bowen2022}. The presence and properties of these fluctuations are in quantitative agreement with solutions to the linear kinetic dispersion relation by {\tt ALPS}, whose input VDFs were taken directly from the simulation output -- a testament to the non-trivial relevance of linear physics in a nonlinear turbulent environment. We have additionally shown how the helicity barrier affects the heating and stability of alphas and minor ions, with heavier masses attaining larger perpendicular temperatures and larger temperature anisotropies (with $T_{\perp,i}>T_{\parallel,i}$). 
Minor ions undergo significant perpendicular heating before protons do. This is because the heavier masses (slower Larmor frequencies) of the minor ions put them into resonance with larger-amplitude oblique-PCWs at smaller $|w_\parallel|$, where the ions are more abundant. These resonance contours are also steeper in the $w_\parallel$-$w_\perp$ plane for heavier species, enabling those ions to climb to extreme perpendicular temperatures.

To demonstrate that the extreme heating of minor ions is a consequence of the turbulence imbalance, we have compared the measured VDFs, perpendicular heating rates, and temperatures with those from an otherwise equivalent simulation of balanced turbulence. This balanced simulation exhibits ion heating that is easily distinguishable from the imbalanced case, and is consistent with theoretical predictions for, and prior hybrid-kinetic simulations of, stochastic ion heating \citep{Chandran2013,KleinChandran2016,Arzamasskiy2019,Cerri2021}.

Both ion cyclotron heating and stochastic heating show dependencies on charge and mass that generally favor the heavier species. Identification of these mechanisms is most apparent in the VDFs. For example, the ion VDFs in the balanced simulation diffuse along $w_{\perp}$ to produce flat-top distributions, a defining feature of  stochastic heating \citep{KleinChandran2016}. They also exhibit a symmetric set of wings in the parallel VDF that correspond to Landau damped forward- and backward-propagating kinetic Alfv\'en waves. In the imbalanced simulation, ion cyclotron heating produces extreme perpendicular ion temperatures, with VDFs that follow the oblique PCW resonance contours to large perpendicular velocities. Landau damping of kinetic Alfv\'en waves occurs primarily in the direction of imbalance, leading to a relatively small amount of parallel heating and flattening of the VDF at the Landau resonance $w_\parallel \approx -v_{\rm A}$. There is also a contribution to parallel heating from ions diffusing along the ends of the cyclotron resonance contours, which follow an arc in $w_{\parallel}$-$w_\perp$ space. This diffusion then leads to the tips of the arcs reaching the Landau resonance and undergoing further parallel heating, even for O$^{5+}$, a species that is initially far from the Landau resonance because of its larger mass. While the heating remains preferentially perpendicular, the VDFs become highly arced towards the parallel direction.

The VDFs from our simulations produce qualitatively recognizable signatures, which if seen in solar wind data could be taken as evidence for the presence of cyclotron heating and/or stochastic heating. A similar diffusion of proton VDFs along cyclotron resonances has already been identified and used as a signature of ion-cyclotron heating in the fast solar wind, as measured by PSP \citep{Bowen2020b,Bowen2022,Bowen2024}. If our predicted arced alpha and minor-ion VDFs were to be observed in future solar-wind data and seen to correlate with turbulence imbalance, a steep transition range in the electromagnetic spectrum just above $k_\perp\rho_{\rm p}\sim 1$, and a large ratio of proton-to-electron heating, then this would constitute strong evidence that the helicity barrier is active and important for the evolution of solar wind. These features could in principle be seen in alpha VDFs measured by PSP or minor-ion VDFs measured by the Heavy Ion Sensor on SO \citep{Livi2023}. Finally, the simultaneous heating of protons, alphas, and O$^{5+}$ ions in our imbalanced simulation give perpendicular temperatures that scale with species mass in a manner consistent with an empirical power-law fit by \citet{Tracy2016} to measurements taken at 1~au by ACE/SWICS of collisionally young solar-wind plasma. If the heating of this plasma is frozen in from its origin close to the Sun, then this agreement provides indirect evidence that the turbulence imbalance in the near-Sun solar wind impacts the plasma heating.

Forthcoming work will demonstrate how the differential energization of minor ions depends on the plasma beta and the degree of imbalance, with the goal of incorporating this heating into a model for the global evolution of the solar wind.

\vspace{1cm}
\noindent This work benefited from useful conversations with Benjamin Chandran, Christopher Chen, Mihailo Martinovi\'c, Romain Meyrand, and Evan Yerger. M.F.Z. and M.W.K. were supported by the National Aeronautics and Space Administration (NASA) under Grant No.~80NSSC24K0171 issued through the Heliophysics, Theory, Modeling and Simulation Program. Support for J.S.~was provided by Rutherford Discovery Fellowship RDF-U001804 and Marsden Fund grant UOO1727, which are managed through the Royal Society Te Ap\=arangi. K.G.K. was supported by NASA grant ~80NSSC19K0912 from the Heliophysics Early Career Investigators Program. This work is part of the Frontera computing project at the Texas Advanced Computing Center under allocation number AST20010; it also made extensive use of the Perseus cluster at the PICSciE-OIT TIGRESS High Performance Computing Center and Visualization Laboratory at Princeton University, as well as High Performance Computing resources supported by the University of Arizona Research Technologies department. The authors thank the Kavli Institute for Theoretical Physics (KITP) for its hospitality during the completion of this work; KITP is supported in part by the National Science Foundation under Award No.~PHY-2309135.

%\bibliography{main.bib}
%\bibliographystyle{aasjournal.bst}

\end{document}